\documentclass[10pt,letterpaper,twocolumn]{article} 

\usepackage{ol2}
\usepackage{amsmath}

\newcommand{\ket}[1]{|#1\rangle}
\newcommand{\bra}[1]{\langle#1|}

\begin{document}

\twocolumn[ 

\title{Index of refraction engineering in five level DIGS atoms}


\author{Steven A. Sagona-Stophel$^{1}$, James Owen Weatherall,$^{2}$ and Christopher P. Search$^{1,*}$}

\address{
$^1$Department of Physics and Engineering Physics, Stevens Institute of Technology, \\ Castle Point on Hudson, Hoboken, NJ 07030, USA\\
$^2$Department of Logic and Philosophy of Science, UC-Irvine \\ 3151 Social Science Plaza A, Irvine, CA 92697, USA \\
$^*$Corresponding author: christopher.search@stevens.edu
}

\begin{abstract} We present a five level atomic system in which the index of refraction of a probe laser can be enhanced or reduced below unity with vanishing absorption in the region between pairs of absorption and gain lines formed by dressing of the atoms with a control laser and RF/microwave fields. By weak incoherent pumping of population into a single metastable state, one can create several narrow amplifying resonances. At frequencies between these gain lines and additional absorption lines, there exist regions of vanishing absorption but resonantly enhanced index of refraction. In $Rb$ vapors with density $N$ in units of $cm^{-3}$, we predict an index of refraction up to $n \approx \sqrt{1+1.2\times10^{-14}N}$ for the D1 line, which is more than an order of magnitude larger than other proposals for index of refraction enhancement. Furthermore, the index can be readily reduced below $1$ by simply changing the sign of the probe or RF field detunings. This enhancement is robust with respect to homogeneous and inhomogeneous broadening.
\end{abstract}


 ] 

Since the wavelength of light in a medium is $\lambda=\lambda_{\text{vac}}/n$, a large refractive index will result in a reduced wavelength relative to the vacuum.  A shorter wavelength, meanwhile, corresponds to enhanced imaging resolution for microscopes and other optical imaging equipment including lithography.  For normal media, however, indices of refraction that differ significantly from the vacuum value of $n=1$ only occur at frequencies with large absorption. This is most obvious in the vicinity of an atomic resonance where the maximum value of the real part of the linear susceptibility, $\Re[\chi]$, occurs at a detuning from resonance equal to the line width where the imaginary part, $\Im[\chi]$, which is proportional to the absorption, has the same magnitude.

In 1991, Scully showed that one can use quantum interference to engineer enhanced indices of refraction with simultaneously vanishing absorption \cite{Scully-EIR}. The essential idea is to use destructive interference resulting from ground state coherence in $\Lambda$ atoms to cancel the optical absorption near an atomic resonance. The original proposal was tested experimentally \cite{Zibrov+etal} and yielded a shift in the index ($\Delta n=n-1$) for $Rb$ vapors of $\Delta n=10^{-4}$, about ten thousand times larger than the background value at the experimental atomic density of $N=10^{12}cm^{-3}$. A more recent proposal uses a pair of closely spaced Raman transitions to create an absorption-gain line doublet with a region of vanishing absorption but enhanced index of refraction between the lines \cite{Yavuz}, which later yielded experimentally index shifts in Rb vapors of $\Delta n=10^{-6}$ at $N=2.4\times 10^{12}cm^{-3}$ \cite{Proite}.

Here we show how a five level atomic system utilizing a single resonant control laser and two RF/microwave fields known as DIGS (Dressed Interacting Ground States) \cite{Weatherall+Search2} can be used to produce not only enhanced indices of refraction but also reduced indices ($0<n<1$) in a tunable spectral region. Unlike Ref. \cite{Yavuz}, the fields drive near resonant transitions rather than far off-resonant Raman transitions resulting in a much larger effect. In comparison to the pumped $\Lambda$ system of Refs. \cite{Scully-EIR, Lukin}, only very weak pumping to a metastable ground state (with negligible excited state population) is necessary thereby minimizing energy loss due to spontaneous emission\cite{Lukin}.  Moreover the enhanced index persists even with significant decoherence between ground states. The predicted index shift with DIGS is up to three orders of magnitude larger than Refs. \cite{Yavuz} and \cite{Proite} and an order of magnitude larger than Refs. \cite{Scully-EIR} and \cite{Zibrov+etal}.

A DIGS atom consists of an excited state $\ket{a}$ coupled to two pairs of hyperfine ground states, $\{\ket{b},\ket{b'}\}$ and $\{\ket{c},\ket{c'}\}$ \cite{Weatherall+Search2}. The ground states are driven by control RF/microwave fields of frequency $\nu_b$ and $\nu_c$ and associated Rabi frequencies $\Omega_b$ and $\Omega_c$.  $\ket{c}$ is strongly coupled to $\ket{a}$ by a resonant control laser of frequency $\nu_{\mu}$ and Rabi frequency $\Omega_{\mu}$, and we consider the optical response as measured by a weak probe of frequency $\nu_p$ (and Rabi frequency $\Omega_p$) probing the $\ket{a}\leftrightarrow\ket{b}$ transition. The Hamiltonian is
\begin{align}
\mathcal{H}=&\frac{\hbar}{2}\left(\Delta_p\ket{a}\bra{a}+\delta\ket{c}\bra{c}+\Delta_b\ket{b'}\bra{b'}+(\delta+\Delta_c)\ket{c'}\bra{c'}\right.\notag\\
\;\;&\left.-\Omega_{p}\ket{a}\bra{b}-\Omega_{\mu} \ket{a}\bra{c}-\Omega_{b}\ket{b'}\bra{b}
-\Omega_{c}\ket{c'}\bra{c}\right)+\text{h.c.} \nonumber
\end{align}
where $\Delta_p=\omega_a-\omega_b-\nu_p$ is the probe detuning, $\delta=\omega_c+\nu_{\mu}-\omega_b-\nu_p=\Delta_p-\Delta_\mu$ is the two photon detuning, and $\Delta_b=\omega_{b'}-\omega_b-\nu_b$ and $\Delta_c=\omega_{c'}-\omega_c-\nu_c$ are the RF detunings. The probe linear susceptibility is $\chi(\Delta_p)=\frac{2N\wp_{ab}}{\epsilon_0 \mathcal{E}_p}\rho_{ab}(\Delta_p)$, where $N$ is the atomic density, $\wp_{ab}$ is the dipole moment matrix element of the $\ket{a}\leftrightarrow\ket{b}$ transition, and $\Omega_p=2\wp_{ab}\mathcal{E}_p/\hbar$. Here we will work with the reduced susceptibility, $\tilde{\chi}=\frac{\hbar\gamma_{ab}\epsilon_0}{N\wp_{ab}^2}\chi=\frac{2\gamma_{ab}}{\Omega_p}\rho_{ab}$, which is independent of the atomic density. The optical absorption is $\alpha\approx(\pi/\lambda_p)\Im[\chi]$ while in regions of negligible absorption, the index of refraction is $n=(|1+\Re[\chi]|)^{1/2}=(|1+(3N\lambda_p^3/4\pi^2)\Re[\tilde{\chi}]|)^{1/2}$.

We solve for the density matrix coherence element $\rho_{ab}$ including in our equations incoherent pumping $r_j$ for levels $\ket{b}$ and $\ket{c'}$ and decay/decoherence rates $\gamma_{jk}=\gamma_{kj}$ for the density matrix elements $\rho_{jk}=\bra{j}\hat{\rho}\ket{k}$:
$\dot{\rho}_{bb}\propto r_b$, $\dot{\rho}_{c'c'}\propto r_{c'}$, and $\dot{\rho}_{jk}\propto -\gamma_{jk} \rho_{jk}$. We model the decoherence between the ground state subspaces $\{\ket{b},\ket{b'}\}$ and $\{\ket{c},\ket{c'}\}$ with rates $\gamma_{cb}=\gamma_{cb'}=\gamma_C$ and $\gamma_{c'b}=\gamma_{c'b'}=\gamma_{C'}$. Here, for simplicity, we model an open system where all levels decay to outside states but with $\gamma_{aj} \gg \gamma_{kl}$ for $k,l\neq a$ since spontaneous emission from $\ket{a}$ is assumed to be much larger than decay from the lower states. For a weak probe $\Omega_p\ll \gamma_{aa}$ and $\Omega_c \ll \Omega_{\mu}\leq \gamma_{aa}$, the population in the excited state $\ket{a}$ is negligible with the excited state population scaling as $\rho_{aa}\sim (\frac{\Omega_p^2}{4\gamma_{aa}\gamma_{ab}})\frac{r_b}{2\gamma_{bb}}+(\frac{\Omega_\mu}{4\gamma_{aa}})(\frac{\Omega_c}{\Omega_{\mu}})\frac{r_{c'}}{2\gamma_{ac'}(\Omega_c/\Omega_{\mu})^2+\gamma_{c'c'}}$.
The steady state solution of the probe susceptibility for $\Delta_{\mu}=\Delta_c=0$ is $\tilde{\chi}=X_{+}-X_{-}$,
\begin{eqnarray}
X_+&=-\eta\frac{P_B^+(i\epsilon_1+a_+)(i\epsilon_2+a_+)-c^2(P_B^+-P_C^+)} {i\epsilon_2+a_+-(i\eta+a_+)\left((i\epsilon_1+a_+)(i\epsilon_2+a_+)-c^2\right)} \\
X_-&=-\eta\frac{P_B^-(i\epsilon_1+a_-)(i\epsilon_2+a_-)-c^2(P_B^--P_C^-)} {i\epsilon_2+a_--(i\eta+a_-)\left((i\epsilon_1+a_-)(i\epsilon_2+a_-)-c^2\right)}
\end{eqnarray}
where $\eta=\frac{2\gamma_{ab}}{\Omega_{\mu}}$, $c=\frac{\Omega_c}{\Omega_{\mu}}$, $b=\frac{\sqrt{\Omega_b^2+\Delta_b^2}}{\Omega_{\mu}}$,
$\epsilon_1 = \frac{2\gamma_C}{\Omega_{\mu}}$, and $\epsilon_2=\frac{2\gamma_{C'}}{\Omega_{\mu}}$. Finally, the probe detuning is expressed as $a_{\pm}=-2a\mp b$ where $a=\frac{\Delta_p-\Delta_b/2}{\Omega_{\mu}}$. $\chi$ depends on the steady state populations of the states $\ket{c'}$, which to lowest order in $\Omega_c/\Omega_{\mu}$ is given by $\rho_{c'c'}=\frac{r_{c'}\Omega_{\mu}^2}{2\gamma_{c'a}\Omega_c^2+\gamma_{c'c'}\Omega_{\mu}^2}$, and $\ket{b}$, $\rho_{bb}=\frac{r_b(2\gamma_{b'b'}(\gamma_{bb'}^2+\Delta_b^2)+\gamma_{bb'}\Omega_b^2)}{2\gamma_{bb}\gamma_{b'b'}(\gamma_{bb'}^2+\Delta_b^2) +(\gamma_{bb}+\gamma_{b'b'})\gamma_{bb'}\Omega_b^2}$, as well as on the coherence $\rho_{bb'}$, $\rho_{bb'}=\frac{r_b\gamma_{b'b'}(-i\gamma_{bb'}+\Delta_b)\Omega_b}{2\gamma_{bb}\gamma_{b'b'}(\gamma_{bb'}^2+\Delta_b^2) +(\gamma_{bb}+\gamma_{b'b'})\gamma_{bb'}\Omega_b^2}$. Using the dressing angle for the dressed states formed by $\Omega_b$, $\tan2\theta_b=\Omega_b/\Delta_b$, one has  $P_C^+=-\cos^2\theta_b\rho_{c'c'}$, $P_C^-=\sin^2\theta_b\rho_{c'c'}$ for and $P_{B}^{+}=-\cos^2\theta_b\rho_{bb}-\sin\theta_b\cos\theta_b\rho_{bb'}$ and $P_B^{-}=\sin^2\theta_b\rho_{bb}-\sin\theta_b\cos\theta_b\rho_{bb'}$.

\begin{figure}[htb]
\centerline{
\includegraphics[width=0.9\columnwidth,height=2.9in]{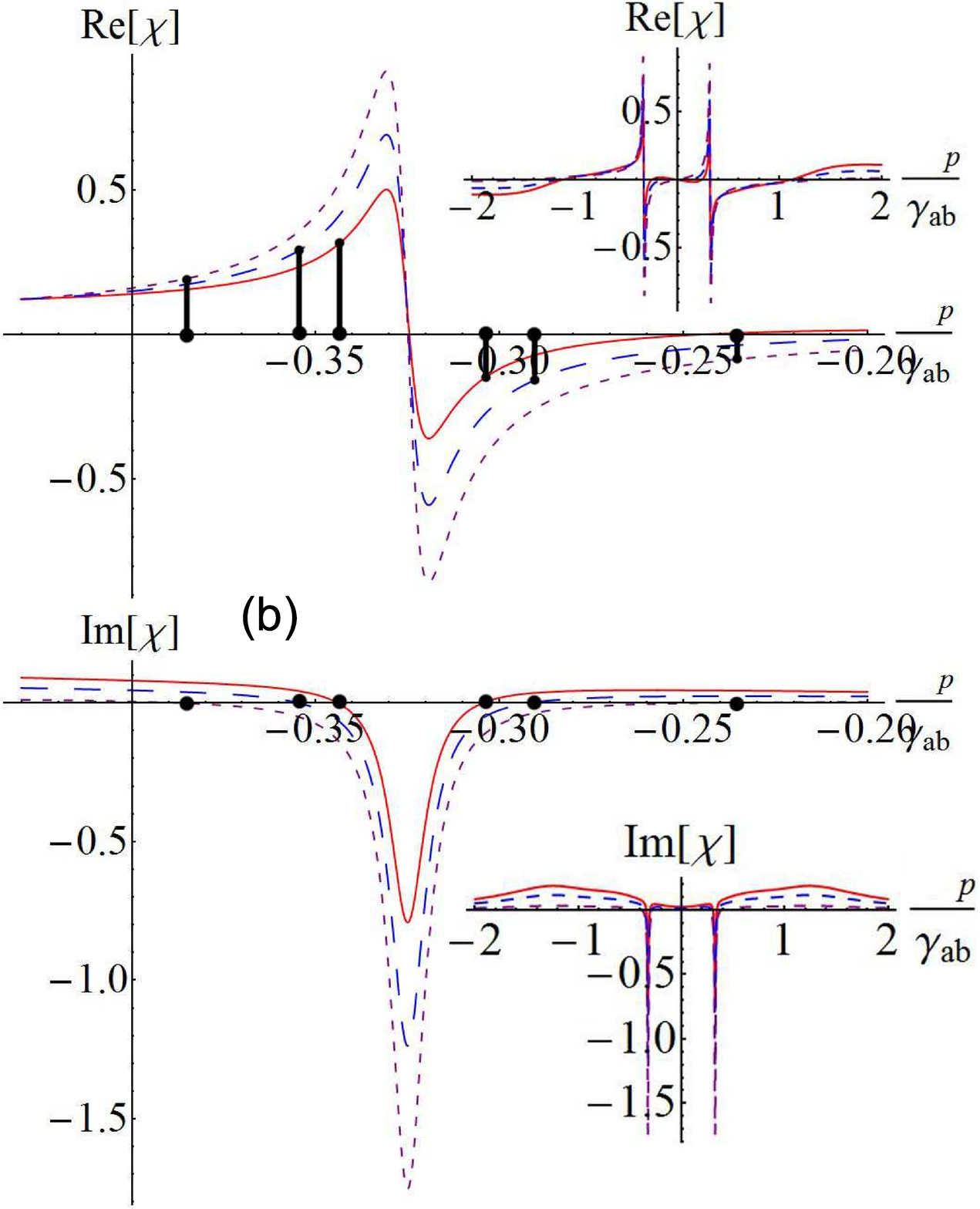}}
 \caption{\label{fig-populations} (a) $\Re[\tilde{\chi}]$ and (b) $\Im[\tilde{\chi}]$ for $\Delta_b=\Delta_c=\Delta_{\mu}=0$ and pumping values $r_b=5\times 10^{-5}$, $r_{c'}=0.023$ (red solid line); $r_b=3\times 10^{-5}$, $r_{c'}=0.03$ (blue dashed line); $r_b=9\times 10^{-6}$, $r_{c'}=0.04$ (purple dotted line). In (b) are the zeros of $\Im[\tilde{\chi}]$ marked with dots located between the absorption lines centered at $\Delta_p=\pm \Omega_\mu/2$ and the narrow gain lines at $\pm \Omega_b/2$. In (a) the locations of the absorption zeros are marked by vertical lines with dots. Here $\Omega_{\mu}=2$, $\Omega_b=0.65$, and $\Omega_c=0.15$ while the decay rates are $\gamma_{aa}=2$, $\gamma_{ab}=\gamma_{ac}=\gamma_{ac'}=1$, $\gamma_{C}=\gamma_{C'}=\gamma_{bb}=\gamma_{b'b'}=\gamma_{cc}=\gamma_{c'c'}=\gamma_{cc'}=\gamma_{bb'}=0.0001$.  Note that all parameters here and in the other figures are measured in units of the bare $\ket{a}\leftrightarrow \ket{b}$ line width, $\gamma_{ab}$. The insets show the same results over a larger range of $\Delta_p$.}
\end{figure}

\begin{figure}[htb]
\centerline{
\includegraphics[width=\columnwidth]{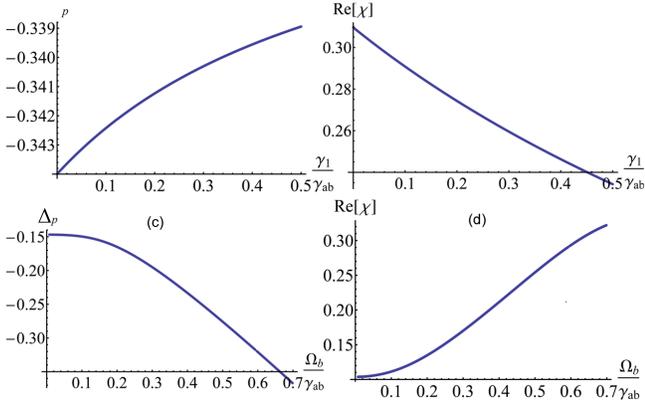}}
 \caption{\label{fig-enhancedIndex} (a) Location of the the absorption zero $\Im[\chi]=0$ in the range $\Delta_p=-\Omega_{\mu}/2$ to $-\Omega_b/2$ and (b) value of $\Re[\tilde{\chi}]$ at the absorption zero as a function of decoherence rate $\gamma_1=\gamma_C=\gamma_{C'}$; (c) Location of the absorption zero $\Im[\chi]=0$ and (d) value of $\Re[\tilde{\chi}]$ at the absorption zero as a function of $\Omega_b$. Pumping is $r_{c'}=0.023$ and $r_b=5\times 10^{-5}$ and all other parameters are the same as in Fig. 1.}
\end{figure}

First we consider the resonant case $\Delta_b=\Delta_c=\Delta_{\mu}=0$. When the pumping exceeds $r_{c'}/r_{b}>(2\gamma_{c'a}\Omega_c^2+\gamma_{C'}\Omega_\mu^2)/((\gamma_{bb}+\gamma_{b'b'})\Omega_{\mu}^2)$, the absorption spectrum consists of two narrow gain lines located at $\Delta_p=\pm \Omega_b/2$ with line width $\Gamma=\gamma_{ab}(\Omega_c/\Omega_{\mu})^2+\gamma_{C'}$ and two broader Autler-Townes absorption lines at $\Delta_p=\pm\Omega_\mu/2$ \cite{Weatherall+Search2} as shown in Fig. 1. In the region between the gain and absorption lines there are two absorption zeros ($\Im[\tilde{\chi}]=0$) where $\Re[\tilde{\chi}] \approx \pm (0.1-0.30)$. For $\Re[\tilde{\chi}]=0.3$ this  corresponds to a $\Delta n=2.6$ at $N=10^{15}cm^{-3}$ and $\lambda_p=800nm$. The magnitude of $\Re[\tilde{\chi}]$ increases linearly with $\Omega_b$ as does the distance of the absorption nulls from $\Delta_p=0$. Most significantly of all is that the absorption nulls are positioned symmetrically  around $\Delta_p=0$ while the values $\Re[\tilde{\chi}]$ at these locations are antisymmetric with respect to $\Delta_p$ with negative $\Re[\tilde{\chi}]$ at positive detunings. Consequently, at negative detunings, $\Delta n>1$ while at positive detunings, $\Delta n<1$. The narrow gain lines are the result of transitions from the excited state $\ket{a_0}=(\Omega_c\ket{a}-\Omega_{\mu}\ket{c'})/\sqrt{\Omega_{\mu}^2+\Omega_{c}^2}$ to the dressed states $\ket{B}=\cos\theta_b\ket{b}+\sin\theta_b\ket{b'}$ and $\ket{B'}=-\sin\theta_b\ket{b}+\cos\theta_b\ket{b'}$. The absorption nulls and $\Re[\tilde{\chi}]$ are hence primarily affected by the decoherence rate $\gamma_{C'}$ and only very weakly indirectly affected by decoherences $\gamma_{cc'}$ and $\gamma_C$. Note in Fig. \ref{fig-enhancedIndex} that even when $\gamma_{C'}\gg\gamma_{ab}(\Omega_c/\Omega_{\mu})^2$, the decoherence does not destroy the absorption nulls but rather pushes them farther from the line centers where  $|\Re[\tilde{\chi}]|$ is smaller.  However, even at ground state decoherence rates equal to $50\%$ of the bare probe line width $\gamma_{ab}$, $\Re[\tilde{\chi}]\approx 0.22$ as seen in Fig. 2(b). Doppler broadening of $\delta$ can be eliminated if a Doppler free geometry with nearly co-propagating lasers is chosen. Nevertheless, as shown in Fig. 3, even though Doppler broadening of $\delta$ suppresses the gain lines and dispersive structure of $\Re[\tilde{\chi}]$ for $\sigma_{\delta}>\Gamma$, the absorption zeros still survive with enhanced nonzero $\Re[\chi]$.
$\chi$ is much less sensitive to Doppler broadening for $\Delta_p$ alone with the absorption nulls and $\Re[\tilde{\chi}]$ experiencing only a small changes up to Doppler widths of $10\gamma_{ab}$.

Next we allow for $\Delta_b\neq 0$ and $r_{c'}/r_{b}\approx (2\gamma_{c'a}\Omega_c^2+\gamma_{C'}\Omega_\mu^2)/((\gamma_{bb}+\gamma_{b'b'})\Omega_{\mu}^2)$ in Fig. 4. At these pumping values the narrow lines are on the border between absorption lines (smaller $r_{c'}/r_{b}$) and gain lines (larger $r_{c'}/r_{b}$). For nonzero $\Delta_b$ the two lines are no longer symmetric due to the unequal distribution of population between $\ket{B}$ and $\ket{B'}$. For increasing positive detuning, $\rho_{B'B'}>\rho_{c'c'}$ transforming the gain line into an absorption line while the other line remains amplifying since $\rho_{BB}<\rho_{c'c'}$. In between the lines is again an absorption null with $\Re[\tilde{\chi}]\approx 0.02-0.05$ corresponding to $\Delta n\approx 0.33-0.73$ at $N=10^{15}cm^{-3}$. Note that by flipping the sign of $\Delta_b$, the populations $\rho_{BB}$ and $\rho_{B'B'}$ are interchanged and the resonances change sign. Simultaneously, $\Re[\tilde{\chi}]$ also changes sign leading to $n<1$. In this case for $\Re[\tilde{\chi}]=-0.02$, $\Delta n=-0.55$ at $N=10^{15}cm^{-3}$.

\begin{figure}[htb]
\centerline{
\includegraphics[width=0.9\columnwidth,height=2in]{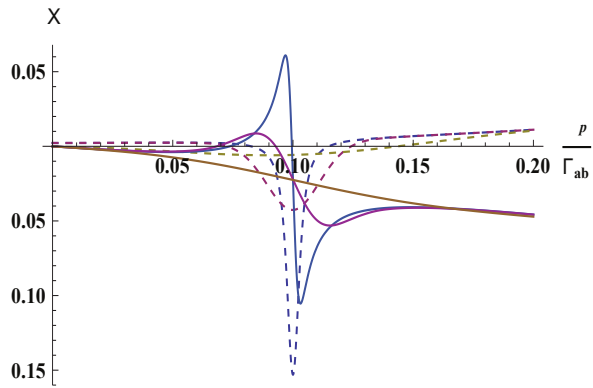}}
 \caption{\label{fig-model} The effect on $\tilde{\chi}$ of Doppler broadening of the two photon detuning $\delta$ for Gaussian distributions of width $\sigma_{\delta}=0.001$ (blue), $\sigma_{\delta}=0.01\approx \Gamma$ (purple), and $\sigma_{\delta}=0.05>\Gamma$ (yellow) in the vicinity of the gain line and absorption null located at $\Delta_p= \Omega_b/2$. Solid lines are $\Re[\chi]$ and dashed lines are $\Im[\chi]$. The other parameters are: $r_b=0.00004$, $r_{c'} =0.0058$, $\Omega_b=0.2$, $\Omega_c=0.1$, $\gamma_{ij}=0.0001$ for $i,j\neq a$, and $\Delta_b=\Delta_c=0$.}
\end{figure}

\begin{figure}[htb]c
\centerline{
\includegraphics[width=0.9\columnwidth,height=2.9in]{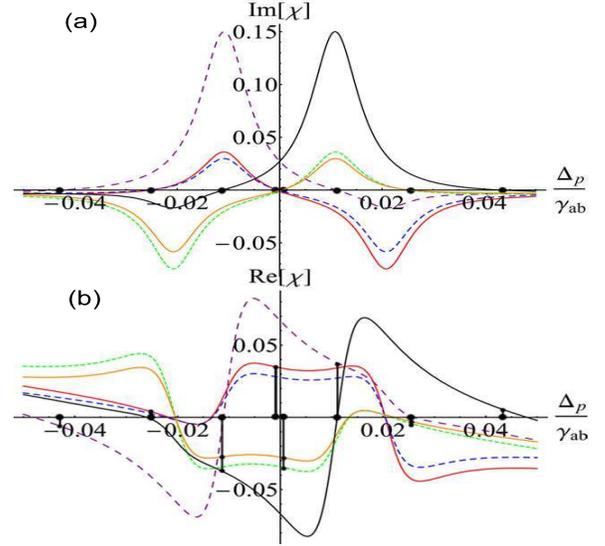}}
 \caption{(a) $\Im[\tilde{\chi}]$ and (b) $\Re[\tilde{\chi}]$ for $\Delta_c=\Delta_{\mu}=0$. Clearly visible in (a) are the zeros of $\Im[\tilde{\chi}]$ marked with dots. The values of $\Re[\chi]$ in (b) where $\Im[\tilde{\chi}]=0$ are marked with vertical lines. Parameters: $r_{c'}=0.0058$, $r_{b}=8.7\times 10^{-5}$, $\Delta_b=0.01$ (red line); $r_{c'}=0.0046$, $r_b=7\times 10^{-5}$, $\Delta_b=0.01$ (blue dashed line); $r_b=9\times 10^{-5}$, $r_{c'}=0.004$, $\Delta_b=0.01$ (purple dashed line); $r_{c'}=0.0046$, $r_b=7\times 10^{-5}$, $\Delta_b=-0.01$ (orange dotted line);  $r_{c'}=0.0058$, $r_{b}=8.7\times 10^{-5}$, $\Delta_b=-0.01$ (green dotted line); $r_b=9\times 10^{-5}$, $r_{c'}=0.004$, $\Delta_b=-0.01$ (black dashed line). All other parameters are identical to Fig. 1.}
\end{figure}

In conclusion, by adding a small amount of population to an additional ground state of 5 level atoms, one can achieve large robust enhancements to the index of refraction with vanishing absorption and negligible energy dissipation from spontaneous emission. The sign of $\Re[\chi]$ can be changed with a change of the sign of the detuning of the probe laser or RF field. Note, however, that this does not lead to negative indices of refraction when $\Re[\chi]<0$ since for $n<0$ a negative magnetic permeability is also required. By contrast for $-1<\Re[\chi]<0$, one has $0<n<1$, and for $\Re[\chi]<-1$ one finds $n>1$ \cite{Fleischhauer}.



\end{document}